\documentclass[]{theotage}
\newdimen\makeboxdimen
\newcommand\makeboxlike[3][l]{%
\setbox0=\hbox{#2}%
\global\makeboxdimen=\wd0%
\setbox1=\hbox{\makebox[\makeboxdimen][#1]{%
\makebox[0pt][#1]{#3}%
}}%
\ht1=\ht0%
\dp1=\dp0%
\box1%
}

\newcommand\like[3][c]{\makeboxlike[#1]{\ensuremath{#2}}{\ensuremath{#3}}}

\newcommand\plaincenter[1]{%
	\mbox{}\hfill#1\hfill\mbox{}%
}

\newcommand\wwrel[1]{\mathrel{\;\;{#1}\;\;}}

\newcommand\wbin[1]{\mathbin{\;{#1}\;}}

\input{ushyphex.tex} 
\usepackage{placeins}

\usepackage{float}
\floatstyle{ruled}
\newfloat{algorithm}{tbp}{loa}
\floatname{algorithm}{Algorithm}

\usepackage{clrscode3e}

\def\EndIf{\End\li\kw{end if} }
\def\EndWhile{\End\li\kw{end while} }

\makeatletter
\renewcommand{\liprint}{\scriptsize{\protected@xdef\@lilabel{\thecodelinenumber}}%
\ifnumberedline{$\thecodelinenumber$}\fi\'\Indent%
}
\makeatother
\renewcommand{\Comment}{\scriptsize \CommentSymbol\ }

\usepackage[american]{wref}

\usepackage{pgfplots}
\pgfplotsset{compat=1.5}
\usepgfplotslibrary{dateplot}

\usepackage{wrapfig}
\def\hacklabeltoalgorithm{
	\makeatletter
	\def\@captype{algorithm}%
	\makeatother
}
\newlength{\originalparindent}
\AtBeginDocument{\setlength{\originalparindent}{\parindent}}

\usepackage{microtype}
\usepackage{url}

\hypersetup{
 			unicode=true, 
            bookmarks=true,
            bookmarksnumbered=true,
            bookmarksopen=true,
            breaklinks=true,
            pdfborder={0 0 0}, 
            colorlinks=false   
}

\begin{document}

\title{Why Is Dual-Pivot Quicksort Fast?}
\author[a]{Sebastian Wild}

\address[a]{Fachbereich Informatik, TU Kaiserslautern\\
  \email{wild@cs.uni-kl.de}}

\maketitle

\begin{abstract}
	I discuss the new dual-pivot Quicksort that is nowadays 
	used to sort arrays of primitive types in Java.
	I sketch theoretical analyses of this algorithm that offer a possible, 
	and in my opinion plausible, explanation why 
	(a) dual-pivot Quicksort is faster than the previously used (classic) Quicksort and 
	(b) why this improvement was not already found much earlier. 
\end{abstract}

\section{Introduction}
Quicksort became popular shortly after its original presentation by Hoare~\cite{Hoare1962} and
many authors contributed variations of and implementation tricks for the basic algorithm.
From a practical point of view, the most notable improvements appear 
in~\cite{Singleton1969,Sedgewick1975,Bentley1993,Musser1997}.\strut

\noindent
\begin{minipage}{.44\linewidth}
\setlength{\parindent}{\originalparindent}
\strut
After the improvements to Quicksort in the 1990's, 
almost all programming libraries used almost identical versions of the algorithm: 
the classic Quicksort implementation had reached calm waters.

It was not until 2009, over a decade later,
that previously unknown Russian developer Vladimir Yaroslavskiy caused a sea change
upon releasing the outcome of his free-time experiments to the public:
a dual-pivot Quicksort implementation 
that clearly outperforms the classic Quicksort in Oracle's Java~6.
The core innovation is the arguably natural ternary partitioning algorithm
given to the right.

Yaroslavskiy's finding was so surprising that
people were initially reluctant to believe him, but
his Quicksort has finally been deployed to millions of devices
with the release of Java~7 in 2011.


How could this substantial improvement to the well-studied Quicksort algorithm
escape the eyes of researchers around the world for nearly 50 years?

\end{minipage}%
\hfill%
\fbox{%
\begin{minipage}{.52\linewidth}
	\vspace*{-2ex}
	\footnotesize
\begin{codebox}
\Procname{~$\proc{DualPivotQuicksort}(A,\id{left},\id{right})$ 
	\;\Comment sort $A[\id{left}..\id{right}]$}
\li \If $\id{right} - \id{left} \ge 1$
\zi \Then 
		\Comment Take outermost elements as pivots (replace by sampling)
\li		$p\gets \like[l]{\max}{\min}\{A[\id{left}],A[\id{right}]\}$ 
\li		$q\gets \max\{A[\id{left}],A[\id{right}]\}$
\li 	$\ell\gets \id{left}+1$;\; $g\gets \id{right}-1$;\; $k\gets \ell$ 
				\label{lin:yaroslavskiy-init-l-g-k}
\li		\While $k\le g$ 
\li 	\Do
			\If $A[k] < p$ \label{lin:yaroslavskiy-comp-1}
\li			\Then
				Swap $A[k]$ and $A[\ell]$;\; \label{lin:yaroslavskiy-swap-1}
				$\ell\gets \ell+1$ \label{lin:yaroslavskiy-l++-1}
\li			\Else \If $A[k] \ge q$ \label{lin:yaroslavskiy-comp-2}
\li				\While $A[g] > q$ and $k<g$ 
\li					\Do 
						$g\gets g-1$ 
				\EndWhile
						\label{lin:yaroslavskiy-comp-3}
\li				Swap $A[k]$ and $A[g]$;\; \label{lin:yaroslavskiy-swap-2}
				$g\gets g-1$ \label{lin:yaroslavskiy-g--}
\li				\If $A[k] < p$ \label{lin:yaroslavskiy-comp-4}
\li				\Then
					Swap $A[k]$ and $A[\ell]$;\; \label{lin:yaroslavskiy-swap-3}
					$\ell\gets \ell+1$ \label{lin:yaroslavskiy-l++-2}
				\EndIf
			\EndIf
\li		$k\gets k+1$ \label{lin:yaroslavskiy-k++}
		\EndWhile \label{lin:yaroslavskiy-end-while}
\li		$\ell\gets \ell-1$;\; $g\gets g+1$
\li		$A[\id{left}] \gets A[\ell]$;\; $A[\ell]\gets p$
			\>\>\>\>\>\>\quad\Comment $p$ to final position
			\label{lin:yaroslavskiy-swap-4}
\li		$A[\id{right}] \gets A[g]$;\: $A[g] \gets q$ 
			\>\>\>\>\>\>\quad\Comment $q$ to final position
			\label{lin:yaroslavskiy-swap-5}
\li		$\proc{DualPivotQuicksort}(A,
				\makeboxlike[c]{$g+1$}{$\id{left}$},
				\makeboxlike[c]{$g+1$}{$\ell-1$}
			)$
\li		$\proc{DualPivotQuicksort}(A,
				\makeboxlike[c]{$g+1$}{$\ell+1$},
				\makeboxlike[c]{$g+1$}{$g-1$}
			)$
\li		$\proc{DualPivotQuicksort}(A,
				\makeboxlike[c]{$g+1$}{$g+1$},
				\makeboxlike[c]{$g+1$}{$\id{right}$}
			)$
	\EndIf
\end{codebox}

	\vspace*{-2ex}
\end{minipage}%
}

For programs as heavily used as library sorting methods, it is advisable to back up 
experimental data with mathematically proven properties.
The latter consider however only a \emph{model} of reality,
which may or may not reflect accurately enough the behavior of actual machines.

The answer to above question is in part a tale of the pitfalls of theoretical models,
so we start with a summary of the mathematical analysis of Quicksort and
the underlying model in \wref{sec:classic-analysis}.
We then briefly discuss the ``memory wall'' metaphor and its implications for Quicksort in \wref{sec:memory-wall},
and finally propose an alternative model in \wref{sec:scanned-elements} 
that offers an explanation for the superiority of Yaroslavskiy's Quicksort.

\section{Analysis of Quicksort}
\label{sec:classic-analysis}

The classical model for the analysis of sorting algorithm considers the 
average number of key comparisons on random permutations.
Quicksort has been extensively studied under this model,
including variations like choosing the pivot as median of a sample~%
\cite{Hoare1962,VanEmden1970VanFun,Sedgewick1975,Martinez2001,Durand2003pseudonine}:
Let $c_n$ denote the expected number of comparisons used by classic Quicksort 
(as given in \cite{Sedgewick1978}), when each
pivot is chosen as median of a sample of $2t+1$ random elements. $c_n$ fulfills
the recurrence
\begin{align}
\label{eq:recurrence-cmps}
		c_n
	&\wwrel=
		n - 1 \wbin+ 
		\sum_{\substack{0\le j_1,j_2 \le n-1\\j_1+j_2=n-1}}
		\dfrac{ \binom{j_1}{t}\binom{j_2}{t} }{ \binom n{2t+1} }
		(c_{j_1} + c_{j_2})
\end{align}
since $n - 1$ comparisons are needed in the first partitioning step, 
and we have two recursive calls of random sizes, 
where the probability to have sizes $j_1$ and $j_2$ is given by the fraction of
binomials (see \cite{MartinezNebelWild2015aofaFullPaper} for details).
This recurrence can be solved asymptotically~\cite{VanEmden1970VanFun,Sedgewick1975} to
\begin{align*}
		c_n 
	&\wwrel\sim
		\frac{1}{H_{2(t+1)} - H_{t+1}} \cdot n\ln n,
\end{align*}
where $H_n=\sum_{i=1}^n 1/i$ is the $n$th harmonic number 
and $f(n)\sim g(n)$ means $\lim_{n\to\infty} f(n)/g(n) = 1$.
The mathematical details are beyond the scope of this abstract, 
but a rather elementary derivation is possible~\cite{Martinez2001}.
Large values of $t$ are impractical; a good compromise in practice is given by the ``ninther'', 
the median of three medians, each chosen from three elements~\cite{Bentley1993}.
This scheme can be analyzed similarly to the above~\cite{Durand2003pseudonine}.

The author generalized \wref[Equation]{eq:recurrence-cmps} to Yaroslavskiy's Quicksort~%
\cite{Wild2012,Wild2012thesis,MartinezNebelWild2015aofaFullPaper}.
Note that unlike for classic Quicksort, 
the comparison count of Yaroslavskiy's partitioning depends on pivot values,
so its expected value has to be computed over the choices for the pivots.
We obtain for tertiles-of-$(3t+2)$
\begin{align}
\label{eq:recurrence-cmps-yaros}
		c_n
	&\wwrel=
		\biggl(\frac53-\frac1{9t+12}\biggr)\cdot n + O(1) \wbin+ 
		\sum_{\substack{0\le j_1,j_2,j_3 \le n-2\\j_1+j_2+j_3=n-2}}
		\dfrac{ \binom{j_1}{t}\binom{j_2}{t}\binom{j_3}{t} }{ \binom n{3t+2} } \cdot
		(c_{j_1} + c_{j_2} + c_{j_3});
\end{align}
with solution
\begin{align*}
		c_n 
	&\wwrel\sim
		\frac{\frac53-\frac1{9t+12}}{H_{3(t+1)} - H_{t+1}} \cdot n\ln n.
\end{align*}

Oracle's Java runtime library previously used classic Quicksort with ninther, and now uses 
Yaroslavskiy's Quicksort with tertiles-of-five;
the average number of comparisons are asymptotically $1.5697 n\ln n$ vs.\ $1.7043 n\ln n$.
According to the comparison model, Yaroslavskiy's algorithm is significantly \emph{worse} 
than classic Quicksort!
Moreover, this result does not change qualitatively if we consider \emph{all}
primitive instructions of a machine instead of only comparisons%
~\cite{Wild2012,MartinezNebelWild2015aofaFullPaper}.
It is thus not surprising that researchers found the use of 
two pivots not promising~\cite{Sedgewick1975,Hennequin1991}.

But if Yaroslavskiy's algorithm actually uses more comparisons and instructions,
how comes it is still faster in practice?
And why was this discrepancy between theory and practice not noticed earlier?
The reason is most likely related to a phenomenon known as the 
``memory wall''~\cite{Wulf1995,McKee2004} or the ``von-Neumann bottleneck''~\cite{Backus1978}:
Processor speed has been growing considerably faster than memory bandwidth for a long time.

\section{The Memory Wall}
\label{sec:memory-wall}

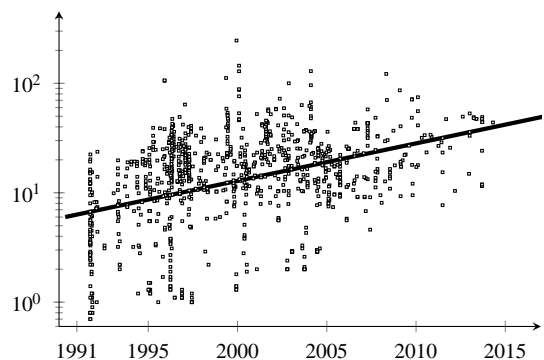
\begin{wrapfigure}[20]{r}{.45\linewidth}
	\vspace*{-5ex}
	\plaincenter{%
	\begin{tikzpicture}[every node/.append style={font=\scriptsize}]
	\def\maxy{455}
	\begin{semilogyaxis}[
	  legend style={ at={(1.1, .97)}, anchor=north west},
	  mark size=.5pt,
	  width=1.1\linewidth,
	  height=0.8\linewidth,
	  ymin=.6,
	  ymax=\maxy,
	  xmin=1990-01-01,
	  xmax=2017-01-01,
	  xtick={1991-01-01,1995-01-01,2000-01-01,2005-01-01,2010-01-01,2015-01-01},
	  axis x line=bottom,
	  axis y line=left,
	  minor tick num=1,
	  date coordinates in=x,
	  xticklabel={\year},
	 ]
		\addplot[only marks,mark=square*,every mark/.append style={fill=white}] 
				table[skip first n=1,x=Date,y={Machine-Balance}] 
					{memory-bandwidth-STREAM.tab} ;
		
		\addplot[mark=none,ultra thick] coordinates 
				{ (1990-05-07,6) (2017-09-22,52) } ; 
	\end{semilogyaxis}
	\end{tikzpicture}%
	}
	\vspace*{-4ex}
	\caption{%
		Development of CPU speed against memory bandwidth over the last 25 years.
		Each point shows one reported result of the 
		STREAM benchmark~\cite{McCalpin1995,McCalpin2007}, with the date on the $x$-axis and
		the machine balance (peak MFLOPS divided by Bandwidth in MW/s in the ``triad'' benchmark)
		on a logarithmic $y$-axis.
		The fat line shows the linear regression (on log-scale).
		Data is taken from \href{http://www.cs.virginia.edu/stream/by_date/Balance.html}{\ttfamily www.cs.virginia.edu/\hspace{0pt}stream/by\_date/Balance.html}.
	}
	\label{fig:memory-cpu-imbalance}
\end{wrapfigure}

Based on the extensive data for the STREAM benchmark~\cite{McCalpin1995,McCalpin2007},
CPU speed has increased with an average annual growth rate of $46\%$ over the last 25 years, 
whereas memory bandwidth,
the amount of data transferable between RAM and CPU in a given amount of time, 
has increased by $37\%$ per year.
Even though one should not be too strict about the exact numbers as they are averages over
very different architectures, a significant increase in \emph{imbalance} is undeniable.
\wref{fig:memory-cpu-imbalance} gives a direct quantitative view of this trend.

If the imbalance between CPU and memory transfer speed continues to grow exponentially,
at some point in the future any further improvements of CPUs will be futile: 
the processor is waiting for data all the time;
we hit a ``memory wall''.

It is debatable if and when this extreme will be reached~\cite{Ertl2001,McKee2004},
and consequences certainly depend on the application.
In any case, however, the (average) relative costs of memory accesses
have increased significantly over the last 25 years.

So when Quicksort with two pivots was first studied, 
researchers correctly concluded that it does not pay off.
But computers have changed since then, and so must our models.

\section{Scanned Elements}
\label{sec:scanned-elements}

Our new cost model for sorting counts the number of \emph{``scanned elements''}.
An element scan is essentially an accesses ``$A[i]$'' to the input array $A$, 
but we count all accesses as one that
use the same index variable $i$ \emph{and} the same value for $i$.
For example, a linear scan over $A$ entails $n$ scanned elements, 
and several interleaved scans 
(with different index variables) cost the traveled distance, 
summed up over all indices, even when the scanned ranges overlap.
We do not distinguish read and write accesses.

We claim that for algorithms built on potentially interleaved sequential scans, 
in particular for classic and dual-pivot Quicksort,
the number of scanned elements is asymptotically proportional 
to the amount of data transfered between CPU and main memory~\cite{MartinezNebelWild2015aofaFullPaper}.

Scanned elements are related to cache misses~\cite{Kushagra2014}, 
but the latter is a machine-dependent quantity, 
whereas the former is a clean, abstract cost measure that is easy to analyze:
One partitioning step of classic Quicksort scans $A$ exactly once, resulting in $n$ scanned
elements.
In Yaroslavskiy's partitioning, indices $k$ and $g$ together scan $A$ once, but index $\ell$ 
scans the leftmost segment a second time.
On average, the latter contains a third of all elements, yielding $\frac43 n$ 
scanned elements in total.

Using these in recurrences \wref{eq:recurrence-cmps} resp.\ \wref{eq:recurrence-cmps-yaros} 
yields $1.5697 n\ln n$ vs.\ $1.4035 n\ln n$ scanned elements;
the Java~7 Quicksort saves $12\%$ of the element scans over the version in
Java~6, which matches the roughly $10\%$ speedup observed in running time studies.

\section{Conclusion}

Memory speed has not fully kept pace with improvements in processing power. 
This growing imbalance forces us to economize on memory accesses 
in algorithms that were almost entirely CPU-bound in the past, 
and calls for new cost models for the analysis of algorithms.
For sorting algorithms that build on sequential scans over their input,
the proposed ``scanned elements'' counts serve as such a model and 
give a good indication of the amount of memory traffic caused by an algorithm.
It is exactly this data traffic where dual-pivot outclasses classic Quicksort,
offering a plausible explanation for its superiority in practice.

\biblio{quicksort-literature}

\end{document}